\documentclass[preprint,aps,amsmath,amssymb,12pt,showpacs,showkeys,superscriptaddress]{revtex4} 

\usepackage{bbm}
\usepackage[dvips]{graphicx}
\usepackage{times}
\usepackage{color}
\usepackage{placeins}
\usepackage{amsmath,amssymb}  
\usepackage{setspace}

\numberwithin{equation}{section}

\newcommand{\e}{\mathbf{e}}
\renewcommand{\r}{\mathbf{r}}

\newcommand{\F}{\mathbf{F}}
\newcommand{\M}{\mathbf{M}}

\newcommand{\q}{\mathbf{q}}
\newcommand{\s}{\mathbf{s}}
\newcommand{\T}{\mathbf{T}}

\renewcommand{\u}{\mathbf{u}}
\renewcommand{\v}{\mathbf{v}}

\renewcommand{\O}{{\mathcal{O}}}
\newcommand{\sign}{{\rm sign}}
\newcommand{\eps}{\varepsilon}

\newcommand{\micron}{{\mu\rm m}}
\renewcommand{\P}{\mathbf{P}}
\newcommand{\G}{\mathbf{\Gamma}}

\newcommand{\R}{\mathcal{R}}
\renewcommand{\L}{\mathbf{L}}
\newcommand{\omegabf}{\boldsymbol{\omega}}
\newcommand{\mubf}{\boldsymbol{\mu}}
\newcommand{\D}{\mathbf{D}}
\newcommand{\K}{\mathbf{K}}
\newcommand{\C}{\mathbf{C}}
\newcommand{\Omegabf}{\mathbf{\Omega}}
\newcommand{\X}{\mathbf{X}}
\newcommand{\Phibf}{\mathbf{\Phi}}

\usepackage{color}

\begin{document}

\title{A three-sphere swimmer for flagellar synchronization\\
\small (provisionally accepted for publication in the New Journal of Physics)}

\author{Katja Polotzek}
\affiliation{Technische Universit{\"a}t Dresden, Department of Mathematics, Dresden, Germany}
\affiliation{Max Planck Institute for the Physics of Complex Systems, Dresden, Germany}
\author{Benjamin M. Friedrich}
\affiliation{Max Planck Institute for the Physics of Complex Systems, Dresden, Germany}
\email{ben@pks.mpg.de}

\date{\today}


\bibliographystyle{apsrev}


\begin{abstract}
In a recent letter (Friedrich \textit{et al.}, Phys. Rev. Lett. \textbf{109} 138102, 2012), 
a minimal model swimmer was proposed that propels itself at low Reynolds numbers by a revolving motion of a pair of spheres.
The motion of the two spheres can synchronize by virtue of a hydrodynamic coupling
that depends on the motion of the swimmer, but is rather independent of direct hydrodynamic interactions.
This novel synchronization mechanism could account for the 
synchronization of a pair of flagella, \textit{e.g.}\ in the green algae \textit{Chlamydomonas}.
Here, we discuss in detail how swimming and synchronization depend on the geometry of the model swimmer
and compute the swimmer design for optimal synchronization.
Our analysis highlights the role of broken symmetries for swimming and synchronization.
\end{abstract}

\maketitle


\section{Introduction}
Hydrodynamics at the scale of a single cell can be counter-intuitive,
as inertia is negligible and propulsion by thrust becomes impossible \cite{Purcell:1977,Happel:hydro,Lauga:2009}.
In this regime of low Reynolds numbers, 
the hydrodynamic equations are symmetric under time-reversal
and net propulsion requires a swimming stroke that breaks time-reversal symmetry \cite{Shapere:1987}.
In pioneering work, Taylor demonstrated theoretically that self-propulsion by purely viscous forces 
is indeed possible when he studied a minimal system of waving elastic sheets
that abstracts from the undulatory bending waves of the flagella of sperm cells \cite{Taylor:1951}.
Since then, simple model swimmers became an indispensable tool
to explore basic principles of swimming at low Reynolds numbers,
both theoretically
\cite{Lighthill:1952,Najafi:2004,Avron:2005,Dreyfus:2005}
and experimentally
\cite{Dreyfus:2005b,Tierno:2008,Leoni:2009}.

Already in his early work \cite{Taylor:1951}, 
Taylor addressed also the striking biological phenomenon of synchronization among beating flagella.
Eukaryotic flagella are long, slender appendages of many cells \cite{Brennen:1977,Vogel:2005}
that beat actively due to the collective activity of molecular motors inside \cite{Alberts:cell}.
The active periodic bending waves of several flagella can synchronize,
as observed \textit{e.g.} for the two flagella of a pair of nearby sperm cells, \cite{Gray:1928,Woolley:2009}, 
or among the hundreds of short flagella on ciliated epithelia in mammalian airways \cite{Sanderson:1981}.
Recently, the green algae \textit{Chlamydomonas},
a unicellular organisms that swims with a pair of flagella like a breast-swimmer \cite{Ruffer:1985a}, 
emerged as an experimental model system for flagellar synchronization 
\cite{Ruffer:1998a,Polin:2009,Goldstein:2009,Goldstein:2011}.
In \textit{Chlamydomonas}, synchronized flagellar beating is required for swimming along a straight path 
\cite{Ruffer:1985a,Polin:2009}.
Coordination of flagellar beating is physiologically important 
for cellular propulsion also in other systems \cite{Brennen:1977,Berg:2004},
as well as for efficient fluid transport \cite{Cartwright:2004,Osterman:2011}.
Taylor suspected that flagellar synchronization is a hydrodynamic effect \cite{Taylor:1951},
although a definite proof is pending to date \cite{Golestanian:2011}.
Again, simple model systems served as a proof of principle that synchronization by hydrodynamic forces is possible,
both in theory
\cite{Vilfan:2006,Niedermayer:2008,Reichert:2005,Gueron:1999,Kim:2004,Uchida:2011}
and experiment
\cite{Kotar:2010,Bruot:2011,Bruot:2012,Leonardo:2012,Lhermerout:2012}.
These model systems demonstrated equally clearly the need for non-reversible phase dynamics and thus broken symmetries 
\cite{Elfring:2009}.
In these models, symmetries were broken either by wall effects \cite{Vilfan:2006},
by additional elastic degrees of freedom \cite{Niedermayer:2008,Reichert:2005}, 
or by evoking phase-dependent driving forces \cite{Uchida:2011,Bruot:2012}.
Here, we study theoretically a novel class of simple swimmers 
that are inspired by the propulsion of \textit{Chlamydomonas} cells 
and comprise a pair of driven spheres instead of two flagella
(a special case was reported in \cite{Friedrich:2012c}, see also \cite{Bennett:2012}).
The two driven spheres of the swimmer can synchronize their motion 
by hydrodynamic forces in a way that is independent of hydrodynamic interactions between them,
but crucially depends on the swimmer's ability to move:
If the two driven spheres are initially out of phase, an imbalance of local torques causes 
the whole swimmer to rotate (and to translate).
This motion imparts different hydrodynamic friction forces on the two spheres, causing one of them to 
slow down its orbit, and the other one to speed up, which ultimately restores synchrony.
Like swimming, synchronization requires that time-reversal symmetry is broken,
which puts certain constraints on the design of the swimmer.

Our model swimmer exemplifies a generic mechanism for hydrodynamic synchronization.
This synchronization mechanism differs in a fundamental way from other mechanisms described previously 
\cite{Vilfan:2006,Niedermayer:2008,Reichert:2005,Gueron:1999,Kim:2004,Uchida:2011,Kotar:2010,Leonardo:2012}
that depended on hydrodynamic interactions.
This hydrodynamic synchronization mechanism suggests a 
specific physical mechanism for flagellar synchronization in \textit{Chlamydomonas}.
Further, we anticipate that this synchronization mechanism could be exploited for the design of novel man-made micro-swimmers.

\textbf{Outline.}
In section 2, we specify the geometry of the model swimmer as well as its equation of motion,
which couples the phase dynamics of its swimming strokes and the swimming motion.
Following \cite{Friedrich:2012c}, 
this equation of motion is derived as a force balance between internal active driving forces and 
generalized hydrodynamic friction forces using the framework of Lagrangian mechanics for dissipative systems.
In section 3, we discuss self-propulsion and self-rotation of the swimmer, 
while section 4 addresses the phenomenon of phase synchronization 
of its two driven spheres in relation to the swimmer's ability to move.

\section{A minimal three-sphere swimmer for flagellar synchronization}

Inspired by a biological micro-swimmer, the biflagellate green algae \textit{Chlamydomonas},
we proposed a minimal swimmer that retains the basic symmetries of a \textit{Chlamydomonas} cell and
is likewise propelled by two `flagellar actuators' \cite{Friedrich:2012c}.
This model swimmer is shown in figure 1 and consists of three spheres,
which are connected by a frictionless planar scaffold.
Two of the three spheres can move along circular trajectories being
connected by lever arms of length $R$ to the corners $\s_i$, $i=1,2$ of an isosceles triangle.
These two spheres have equal radius $a$ and 
mimic the two flagella of \textit{Chlamydomonas}.
The third sphere of radius $b$ plays a role analogous to the body of this cell.
For simplicity, we assume that the swimmer is planar and that the positions $\r_i$ 
of the three spheres all lie in the $xy$-plane.
The position and spatial orientation of each sphere is thus characterized by just two coordinates
$\r_i=(x_i,y_i,0)$ with respect to the $xyz$-laboratory frame
as well as a single rotation angle $\alpha_i$ for rotations around the $z$-axis.
For brevity, $x=x_3$, $y=y_3$, $\alpha=\alpha_3$.
We explicitly note a material frame for the third sphere,
which consists of the orthonormal vectors 
$\e_1=(\cos \alpha,\sin \alpha,0)$, 
$\e_3=(0,0,1)$ and 
$\e_2=\e_3 \times \e_1$.
The geometry of the swimmer is characterized by the dimensions $l$ and $h$ of the
triangular scaffold such that
$\s_i = \r_3 + (-1)^i l \e_1 + h \e_2$, $i=1,2$, see again figure 1.
The circular orbits of the two flagellar spheres are parametrized by respective
flagellar phase angles $\varphi_i$ with
$\r_i=\s_i + R ( -\sin\varphi_i \e_1 + \cos\varphi_i\e_2 )$.
Here, the flagellar phase angles are defined as continuous variables such that 
the number of full cycles of the $i$th sphere after a time $t$ is given by 
$\lfloor (\varphi_i(t) - \varphi_i(0))/(2\pi) \rfloor$, $i=1,2$.
Likewise, the number of full rotations of the $i$th sphere with respect to the laboratory frame
is given by
$\lfloor (\alpha_i(t) - \alpha_i(0))/(2\pi) \rfloor$, $i=1,\ldots,3$,
where we have the identity $\alpha_i=\alpha+\varphi_i$, $i=1,2$, for the first and second sphere.
Throughout the text, positive angular velocities $\dot{\varphi}_i$ indicate counter-clockwise rotations (when viewed along $-\e_3$),
while clockwise rotations correspond to negative angular velocities of the respective sphere.

\subsection{Low Reynolds number hydrodynamics}
\label{sec_hydro}

For a swimming microorganism, inertial forces are usually negligible in comparison to viscous forces,
which is reflected by a small Reynolds number \cite{Happel:hydro,Lauga:2009}
\begin{equation}
\mathrm{Re} = \frac{\rho v l}{\eta}.
\end{equation}
Here, $l$ is a length-scale of the swimmer, $v$ its speed,
and $\eta$ and $\rho$ the viscosity and density of the surrounding fluid.
As an example, we estimate $\mathrm{Re}\sim 10^{-3}$ for the unicellular green algae \textit{Chlamydomonas}
using $v\sim 100\,\micron/\mathrm{s}$ and $l\sim\,10\micron$ \cite{Ruffer:1985a}.

In the following, we will always assume that inertial forces are also negligible for our model swimmer.
The hydrodynamics at low Reynolds numbers is then governed by the Stokes equation,
$\nabla p+\eta\nabla^2\u=0$, 
where $p$, $\u$, $\eta$ denote the pressure field, flow field and viscosity of the fluid, respectively \cite{Happel:hydro}.
Importantly, this equation is linear, which implies a linear relationship between
velocities and resultant forces.
For example, a single sphere of radius $a$ 
that moves with translational velocity $\v$ 
through a viscous fluid will exert a 
hydrodynamic friction force $\F=\gamma\v$ on the fluid,
where $\gamma=6\pi\eta a$ is the Stokes friction coefficient for translational motion \cite{Happel:hydro}.
Likewise, a rotation with angular velocity $\omegabf$ induces
a hydrodynamic friction torque $\T'=\gamma_r \omegabf$
with rotational friction coefficient $\gamma_r=8\pi\eta a^3$.
In the following, 
primed torques will always denotes torques relative to the center $\r$ of the respective sphere.

For an ensemble of $N$ spheres with positions $\r_i$ and radii $a_i$, 
the linearity of the Stokes equation 
implies a similar linear relationship \cite{Happel:hydro,Lauga:2009,Reichert:phd}
\begin{equation}
\label{frictmat}
\P_0 = \G_0 \cdot \dot{\q}_0,
\end{equation}
where $\dot{\q}_0$ is a $6N$-vector that combines the 
translational and rotational velocity components of the $N$ spheres,
$\dot{\q}_0 = (v_{1x},v_{1y},v_{1z},\omega_{1x},\omega_{1y},\omega_{1z},\ldots,\omega_{Nz})$,
while the $6N$-vector $\P_0$ combines the components of the resultant
hydrodynamic friction forces,
$\P_0 = (F_{1x},F_{1y},F_{1z},T'_{1x},T'_{1y},T'_{1z},\ldots,T'_{Nz})$.
Finally, $\G_0$ denotes the $6N\times 6N$ grand friction matrix.
In the limit of large separation $r_{ij}=|\r_i-\r_j|\gg a_i$ between the spheres,
the friction matrix becomes diagonal;
the diagonal is then populated by the common Stokes friction coefficients
for translational and rotational motion of single spheres.
Generally, off-diagonal entries of $\G_0$ characterize hydrodynamic interactions
between the spheres:
Any movement of the $i$th sphere will set the surrounding fluid in motion.
If the $j$th is held fixed in the vicinity, it will slow down this fluid flow and exert
a corresponding friction force on the fluid.

The friction matrix $\G_0$ is always symmetric as a consequence of the Lorentz reciprocal theorem \cite{Happel:hydro}.
Furthermore, $\G_0$ is positive definite, 
as the hydrodynamic dissipation rate $\mathcal{R}_h=\dot{\q}_0^T\cdot\P_0$ 
can never become negative \cite{Happel:hydro}
\begin{equation}
\mathcal{R}_h = \dot{\q}_0^T \cdot \G_0 \cdot \dot{\q}_0 \ge 0.
\end{equation}

The friction matrix $\G_0$ can be computed to arbitrary precision
in $a_j/r_{ij}$ by using \textit{e.g.}\ the method of reflections \cite{Dhont:1996,Reichert:phd}.
The first iteration step of this method accounts for all pairwise interactions between the spheres
and is known as the Rotne-Prager approximation.
The Rotne-Prager approximation generalizes the popular Oseen tensor 
and applies to both translational and rotational motion, see appendix \ref{Rod_Prag_appr}.

\subsection{Over-damped Lagrangian mechanics}

We employ the framework of Lagrangian mechanics for dissipative systems \cite{Goldstein:mechanics,Vilfan:2009} to
derive equations of motions of our model swimmer from simple force balances \cite{Friedrich:2012c}.
The swimmer is characterized by just $5$ degrees of freedom
if its internal constraints are taken into account:
The two phase angles $\varphi_1$ and $\varphi_2$ describe the 
orbiting of the two flagellar spheres, while $x$, $y$ and $\alpha$ 
characterize translations and rotations of the swimmer in the plane of swimming.
We introduce a $5$-component vector of generalized coordinates
\begin{equation}
\q = ( x, y, \alpha, \varphi_1, \varphi_2 )^T.
\end{equation}
The full position vector $\q_0$ for the $N=3$ spheres 
can now be uniquely expressed as a function of the generalized coordinates as $\q_0=\q_0(\q)$.
For each degree of freedom, 
we define a generalized hydrodynamic friction force $P_{h,i}$ conjugate to this degree of freedom
such that the rate of hydrodynamic dissipation can be equivalently written as
\begin{equation}
\mathcal{R}_h = 
\P_h \cdot \dot{\q} = 
P_{h,1}\dot{x} + P_{h,2}\dot{y} + P_{h,3}\dot{\alpha} +
P_{h,4}\dot{\varphi}_1 + P_{h,5}\dot{\varphi}_2,
\end{equation}
where the 5-vector $\P_h=(P_{h,1},\ldots,P_{h,5})$ is simply given by 
$\P_{h}=\L^T\cdot\P_{0}$ 
with a $9\times 5$ transformation matrix $L_{ij}=\partial q_{0,i}/\partial q_j$.
The generalized hydrodynamic friction forces $P_{h,i}$ relate linearly to the generalized velocities
$\dot{q}_j$ 
as
$P_{h,i} = \sum_{j=1}^5 \Gamma_{h,ij} \dot{q}_j$.
Here, we have introduced the generalized $5\times 5$ hydrodynamic friction matrix
$\G_h = \L^T\cdot\G_0\cdot\L$, which is just a contracted version of the grand friction matrix $\G_0$.
Note that $\dot{\q}_0 = \L\cdot\dot{\q}$. 

We further include internal dissipation 
related to the motion of the two driven spheres
with associated dissipation rate $\mathcal{R}_\kappa = \kappa(\dot{\varphi_1}^2 + \dot{\varphi_2}^2)$.
The total dissipation rate is now given by
$\mathcal{R}=\mathcal{R}_h+\mathcal{R}_\kappa$ 
and defines a Rayleigh dissipation function for our mechanical system.
We may write
\begin{equation}
\R =\dot{\q}^T\cdot\G\cdot\dot{\q},
\end{equation}
where for notational convenience $\G$ is defined as $\G=\G_h+\G_\kappa$
with a $5 \times 5$ internal friction matrix $\G_\kappa$, 
whose only non-zero entries are $\Gamma_{\kappa,44} = \Gamma_{\kappa,55} = \kappa$.
The Rayleigh dissipation function is related to the 
generalized friction forces $P_i$ by $2P_i = \partial \R / \partial q_i$, $i = 1,\ldots,5$.
Explicitly, these generalized friction forces read
\begin{align}
& P_1 = F_{1,x} + F_{2,x} + F_{3,x}, \quad 
P_2 = F_{1,y} + F_{2,y} + F_{3,y}, \\
& P_3 = \left[ (\r_1-\r_3) \times \F_1 + (\r_2-\r_3) \times \F_2 \right]_z   +T'_{1,z}+T'_{2,z}+T'_{3,z}, \\
& P_{j+3} = \F_j \cdot (\partial\r_j/\partial\varphi_j) +T'_{j,z} + \kappa \dot{\varphi}_j,\quad j=1,2.
\end{align}
Note that $P_1$ and $P_2$ simply denote the $x$- and $y$-component of the total
hydrodynamic friction force exerted by the swimmer on the fluid.
Likewise, $P_3$ denotes the $z$-component of the total torque exerted by 
the swimmer on the fluid (with respect to $\r_3$).
For free swimming, the swimmer is free from external forces and torques,
which implies a force balance, $P_1=P_2=0$
as well as a torque balance $P_3=0$.

The two flagellar spheres are actuated by generalized active driving forces $Q_4$ and $Q_5$ 
(with have units of a torque).
For simplicity, we assume these generalized forces to be constant, independent of the flagellar phase,
$Q_4=m_1$ and $Q_5=m_2$.
The active driving forces can be re-written as potential forces:
The energy for active swimming is provided by a fuel reservoir with internal energy $U$ such that $-\dot{U}=\mathcal{R}$.
The simplest choice 
$U = -m_1 \varphi_1 -m_2 \varphi_2$ 
defines generalized potential forces $Q_i = -\partial U / \partial q_i$, $i = 1,\ldots,5$,
and recovers the above definition with phase-independent $Q_4$ and $Q_5$. 
A case of phase-dependent driving forces was recently studied numerically in \cite{Bennett:2012}.
If inertial forces can be neglected as in our case and the swimmer is free from external forces and torques,
we obtain a balance of generalized forces, $P_i=Q_i$, $i=1,\ldots,5$,
where $Q_1=Q_2=0$ and $Q_3=0$.
In matrix notation, this force balance reads for this case for free swimming
\begin{equation}
\label{eq_balance}
(0,0,0,m_1,m_2)^T = \G\cdot\dot{\q}. 
\end{equation}

\subsection{Equations of motion} 
\label{sec_eom}

\textit{Free swimming.} 
The balance of generalized forces stated in eq.~(\ref{eq_balance}) 
enables us to self-consistently solve for the dynamics $\dot{\q}$ of the swimmer.
It is convenient to introduce a block matrix notation for the generalized friction matrix
\begin{equation}
 \label{gamma_blocks}
 \G = \begin{pmatrix} \K & \C \\ \C^T & \Omegabf \end{pmatrix};
\end{equation}
here $\K$, $\C$ and $\Omegabf$ have dimension $3 \times 3$, $3 \times 2$ and $2 \times 2$, respectively. 
Similarly, we write $\q=(\X,\Phibf)$ with $\X=(x,y,\alpha)$ and $\Phibf=(\varphi_1,\varphi_2)$.
From equation (\ref{eq_balance}), we find
\begin{align}
\label{eq_free1}
\dot{\X} &= -\K^{-1}\C\dot{\Phibf}, \\
\label{eq_free2}
\dot{\Phibf} &= (\Omegabf-\C^T\K^{-1}\C)^{-1}(m_1,m_2)^T.
\end{align}
The first equation (\ref{eq_free1}) states that the motion $\dot{\X}$ 
of the swimmer is fully determined by its phase dynamics $\dot{\Phibf}$.
The phase dynamics $\dot{\Phibf}$ in turn depends only on $\Phibf$,
but not on the position and orientation of the swimmer as encoded by $\X$.
In fact, the friction matrix $\G$ is independent of the position of the swimmer.
Furthermore, a rotation of the swimmer around $\e_3$ 
with rotation matrix $\D$ transforms $\G$ according to
$\K\rightarrow \D^{-1}\K\D$, $\C\rightarrow \D^{-1}\C$, $\Omegabf\rightarrow\Omegabf$.
In particular, $\Omegabf-\C^T\K^{-1}\C$ does not change under such a rotation.

\textit{Clamped third sphere.} 
If the third sphere is constrained from translating and rotating, we naturally have $\dot{\X}=0$.
In this case, a generalized constraining force $\C\dot{\Phibf}$ must be applied at $\r_3$.
The phase dynamics is then given by
\begin{equation}
\label{eq_clamped}
\dot{\Phibf}=\Omegabf^{-1} (m_1,m_2)^T.
\end{equation}

Both equation (\ref{eq_free2}) and equation (\ref{eq_clamped}) 
describe a pair of coupled phase oscillators \cite{Pikovsky:synchronization} and can be recast in the form
\begin{equation}
\dot{\varphi}_i = m_i/\kappa + H_i(\varphi_1,\varphi_2), \quad i=1,2.
\end{equation}
Importantly, the coupling functions $H_i$ are different in the case of free-swimming and
the case of a clamped third sphere.
In the case, where the third sphere is clamped, the coupling functions $H_i$ comprise only direct hydrodynamic interactions 
between the spheres.
In the case of free swimming, the coupling functions $H_i$ implicitly account also for the motion of the swimmer
and the associated hydrodynamic friction forces acting on the two flagellar spheres.

\section{Swimming by hydrodynamic interactions}

\subsection{Swimming forward or backward}

We study the case of exactly opposite driving torques $m_2 = -m_1$,
which gives rise to a counter-rotation motion of the first and second sphere, 
similar to the mirror-symmetric beat of the two flagella of \textit{Chlamydomonas}. 
In this case, there exists a limit cycle of the phase dynamics 
with $\delta=\varphi_1+\varphi_2=0$, which corresponds to in-phase motion of the two spheres.
In section \ref{sec_sync}, we show that this limit cycle is stable for $\omega_0 h<0$ and unstable for $\omega_0 h>0$.
Here, we introduced the (signed) angular frequency $\omega_0=m_1/\kappa$.
The corresponding period is $T=2\pi/\omega_0$.

To gain analytical insight, we resort to a limit of small spheres and short lever arms lengths.
We introduce a small expansion parameter $\eps=a/l$ and assume $b/l$ and $R/l$ to be of order $\O(\eps)$.
We further assume that $\kappa/(\eta l^3)$ is of order unity,
which implies that internal dissipation dominates hydrodynamic dissipation.
With these assumptions, the force balance equation (\ref{eq_balance}) is only weakly non-linear,
and a perturbation calculation for small $\eps$ becomes feasible.
Details of this calculation can be found in appendix \ref{perturb_calc}.

For mirror-symmetric propulsion with $\delta=0$, 
the swimmer will move along a straight path with velocity $v=\dot{\r}_3\cdot\e_2$ and zero rotation $\dot{\alpha}=0$.

Remarkably, even for zero lever arm length $R=0$, 
when the first and second sphere spin only around their respective centers,
the swimmer is able to move forward by harnessing hydrodynamic interactions
that couple to rotational motion, see also \cite{Or:2009} for a similar swimmer design.
For the swimming speed, we find in this case
\begin{equation}
\lim_{R\rightarrow 0} v = v_0,\ 
\text{ where }
v_0 = \omega_0 a \frac{a}{2 a + b}
   \left(
        \frac{a^2}{2 l^2}
      + \frac{2 ab l}{(l^2+h^2)^{3/2}}
   \right) + \O(\eps^4).             
\end{equation}
Note that net propulsion is possible even in the absence of the third sphere, $b=0$,
which provides a simple design of a two-sphere-swimmer.

In the general case of a finite lever arm length $R>0$, 
the swimmer will periodically wiggle back-{\&}-forth
to leading order in $\varepsilon$
\begin{align}
\label{eq_yosc}
v = \omega_0 R \frac{2a}{2 a + b}\sin{\varphi_1} + \O(\eps^2).
\end{align}
Net propulsion is a higher-order effect with time-averaged velocity $\langle v\rangle=\O(\eps^3)$
that requires hydrodynamic interactions \cite{Happel:hydro,Vilfan:2009}.
In the limit of small lever arm lengths, $R\ll l$, we find
\begin{align}
\label{eq_quadratic_law}
\langle v \rangle - v_0 \sim R^2 + \O(R^3),
\end{align}
where $v_0=\lim_{R\rightarrow 0}\langle v\rangle$.
Eq.~(\ref{eq_quadratic_law}) exemplifies 
the basic fact that the net swimming speed,
which a swimmer can achieve by periodic body shape changes,
generically scales with the square of the amplitude of its swimming stroke.
This ``quadratic law of low Reynolds number propulsion'' holds for other model swimmers 
\cite{Lighthill:1952,Najafi:2004,Dreyfus:2005}
and was formally proven by Shapere and Wilczek \cite{Shapere:1987}
by generalizing Purcell's scallop theorem \cite{Purcell:1977}.

While non-zero net propulsion is a generic consequence of a 
non-reciprocal swimming stroke that breaks time-reversal symmetry,
the actual sign of the velocity can depend on the geometry of the swimmer \cite{Becker:2003}.
In our case, $\langle v\rangle$ can switch sign as a function of $b/a$ and $h/l$, see figure 2.
As a consequence, our swimmer may either swim forward or backward for $\omega_0>0$,
\begin{align}
\label{eq_vavg}
\langle v \rangle 
&=v_0 -a b \omega_0 R^2 \frac{3}{\sqrt{2}} \frac{(6-\sqrt{2})a-2 b}{8(2a+b)^2 l^2}+\O(\eps^4)
& \text{ for }h=l, \\
\langle v \rangle
&=v_0 +a b \omega_0 R^2 \frac{3a+12b}{8(2a+b)^2 l^2}+\O(\eps^4)
& \text{ for }h=0.
\end{align}
This surprising behavior 
can be related to two opposing mechanisms that contribute to net propulsion
in the case of in-phase beating with $\delta=0$:
The instantaneous swimming speed $v=f/g$ can be computed
from the friction coefficient $g$ associated with towing the swimmer along the $\e_2$-direction,
and a force $-f\e_2$ that equals the force that had to be applied at $\r_3$ to prevent the swimmer from moving.
(If $\alpha=0$, $g=\K_{22}$ and $(0,f,0)^T=-\C\dot{\Phibf}$.)
The force $f$ oscillates during each swimming stroke,
$f=12\pi\eta a R \omega_0 \sin\varphi_1+\O(\eps^3)$,
which is tightly linked to the back-{\&}-forth motion of the swimmer as characterized by equation (\ref{eq_yosc}).
Now, hydrodynamic interactions between the spheres result in higher-order corrections to $f$ 
and imply in particular $\langle f\rangle >0$ for $\omega_0>0$.
This can be seen as follows: 
When the spheres are close together (a configuration for which $f<0$ if $\omega_0>0$),
hydrodynamic interactions are strongest and \emph{reduce} the magnitude of $f$.
Averaging over one phase cycle thus leaves a positive net contribution.
This is the first mechanism, which contributes to forward swimming for $\omega_0>0$.
At the same time, the friction coefficient $g$ depends on the phase $\varphi_1$:
it is largest, when the spheres are furthest apart, and smallest, when they are close together.
The oscillations in the friction coefficient $g$ 
can resonate with the oscillations of the force $f$ 
and contribute to backward swimming.
Depending on the geometry of the swimmer, either mechanism may dominate
resulting in either 
$\langle v\rangle>0$ or
$\langle v\rangle<0$, see figure 2.

\subsection{Asynchronous beating results in rotational motion}

In the case of a phase-difference between the two driven spheres, $\delta=\varphi_1+\varphi_2\neq 0$,
mirror-symmetry with respect to the central axis of the swimmer is broken 
and the swimmer can rotate with $\dot{\alpha}\neq 0$, where $\alpha=\alpha_3$ for brevity.
These rotations can feed back on the phase dynamics of the two spheres
and cause them to synchronize as discussed in the next section.
In this subsection, we want to elucidate how asynchrony causes the swimmer to rotate,
without further complications arising from the feedback on the phase dynamics.
To this aim, we will temporarily 
prescribe the phase dynamics as $\varphi_1=\omega_0 t$, $\varphi_2=-\omega_0t+\delta$.
We then find that the swimmer ``rocks'' in a periodic fashion,
\begin{equation}
 \dot{\alpha} = - \frac{\omega_0 R \sin(\delta/2)} {(2 a + b) l^2 + b h^2} 
                     \left[ (2 a + b) l \cos(\varphi_1-\delta/2) 
                             - b h           \sin(\varphi_1-\delta/2)
                      \right] + \O(\eps^2).
\end{equation}
Figure 3 further shows the net rotation rate, averaged over one beat cycle.
In the special case, where $l=h$, $b=a$, we have
\begin{align}
\label{eq_alphaavg}
\langle \dot{\alpha} \rangle =
\omega_0 \frac{a R^2}{(8l)^3}(19-8\sqrt{2})\,\sin(\delta) + \O(\eps^4).
\end{align}
A qualitatively similar behavior is found for sideways translation motion with velocity $\dot{\r}_3\cdot\e_1$.

\textit{Role of symmetries.}
The net rotation rate $\langle\dot{\alpha}\rangle$ is always 
an odd function of the phase difference $\delta$.
This behavior is a direct consequence of the symmetries of the swimmer:
A reflection at the central symmetry axis of the swimmer 
maps the swimmer onto a copy of itself, but with phase difference $\delta\rightarrow -\delta$.
Since a reflection changes the sign of $\dot{\alpha}$, the assertion follows.
As a corollary, the swimmer will show zero net rotation for $\delta=\pi$.
In a similar way, $\langle\dot{\alpha}\rangle$ is an odd function also of the position $h$ of the third sphere.
To see this, consider a rotation that 
rotates the swimmer in the $xy$-plane by an angle $\pi$ around the point $(\s_1+\s_2)/2$.
This rotation maps the swimmer onto a swimmer for which both $h$ and $\omega_0$ have the opposite sign.
Of course, $\dot{\alpha}$ does not change under this operation,
which implies that $\langle\dot{\alpha}\rangle$ is an even function of $\omega_0h$.
Since $\omega_0$ sets the time-scale of the problem, we have $\dot{\alpha}\sim\omega_0$
and the assertion follows.
As a particular case, the swimmer will show zero net rotation if $h=0$.
In the next section, we will encounter a similar behavior for the synchronization strength.

\section{Synchronization depends on motion}
\label{sec_sync}

In our model, the phase velocity of each of the two driven spheres, $\dot{\varphi}_i$,
is determined by a balance of an active driving force $Q_{i+3}$ and 
a generalized hydrodynamic friction force $P_{i+3}$, $i=1,2$.
The hydrodynamic friction forces $P_i$ are affected by the instantaneous motion of the swimmer,
which indirectly results in a coupling between the two driven spheres, see section \ref{sec_eom}.
To characterize the resultant phase dynamics,
we resort to a stroboscopic description at times $t_n$ that mark 
the completion of a full cycle of the first sphere,
such that $\varphi_1(t_n)=2\pi n$ with integer $n$.
We introduce a Poincar{\'e} return map $\Lambda(\delta)$
that measures the change of the phase difference $\delta=\varphi_1+\varphi_2$
after a full cycle of the first sphere,
$\Lambda(\delta(t_0))=\delta(t_1)-\delta(t_0)$.
To leading order, we find \cite{Friedrich:2012c}
\begin{equation}
\Lambda(\delta) = - \lambda \sin(\delta) + \O(\eps^6).
\end{equation}
Here, $\lambda=-d \Lambda/d \delta|_{\delta = 0}$ denotes a dimensionless synchronization strength \cite{Adler:1946}:
The limit cycle of in-phase synchronized phase dynamics with $\delta=0$
will be exponentially stable with Lyapunov exponent $\ln(1-\lambda)$ whenever $\lambda>0$.
For free swimming, we find $\lambda>0$ provided $\omega_0 h<0$
\begin{equation}
\label{eq_lambda}
\lambda = - \sign (\omega_0 h)\, 
\frac{2 b l |h| \eta}{\kappa}
\left( 
\frac{3 \pi a R^2}{(2 a+b) l^2 + b h^2}
\right)^2 
+\O(\eps^6),
\end{equation}
see also figure 4D.
Expression (\ref{eq_lambda}) for the synchronization strength $\lambda$ has a single global maximum 
as a function of the size $b$ and position $h$ of the third sphere, 
which is characterized by $b=a$ and $h=l$.
Thus, the swimmer design shown in figure 1 is optimal for synchronization in the case of free swimming.
Interestingly, hydrodynamic interactions contribute only to higher order 
to the synchronization strength $\lambda$ \cite{Friedrich:2012c}.
Instead, synchronization strongly depends on the swimmers ability to move.
For our particular swimmer design, strong synchronization with $\lambda=\O(\eps^5)$
requires both translational and rotational motion.
As a means of illustration, consider the case, 
where translation of the swimmer is constrained by an external force,
while the swimmer can still rotate freely. 
We then find
$\lambda=\O(\eps^6)$, more specifically
\begin{equation}
\label{eq_lambda_rot}
\lambda = -\sign(\omega_0 h)\,
\frac{l^2|h|\eta}{\kappa}\,
\left[17+5 (h/l)^2 - 8(a/R)^2(3+(h/l)^2)\right]\,
\left(\frac{3\pi aR^2}{4 l(l^2+h^2)}\right)^2
+\O(\eps^7).
\end{equation}          
In this case, hydrodynamic interactions contribute to leading order.
Similarly, if the swimmer can only translate, but not rotate, we obtain
\begin{equation}
\lambda = \sign(\omega_0h)\,
\frac{30ab|h|\eta}{\kappa}\,
\frac{l(l^2-h^2)}{(l^2+h^2)^{3/2}}\,
\left(\frac{3\pi aR^2}{(2a+b)(l^2+h^2)}\right)^2
+\O(\eps^7).
\end{equation}
Finally, if the swimmer is fully clamped and can neither translate nor rotate,
we have $\lambda=\O(\eps^8)$.

\textit{Role of symmetries.}
For any dynamics of the swimmer, we can realize the time-reversed dynamics
by simply reversing the sign of the active driving forces, $m_1\rightarrow -m_1$ and $m_2\rightarrow -m_2$.
Under this mapping, a stable limit cycle of synchronized phase dynamics will become unstable, and vice versa.
Generally, synchronization requires that the phase dynamics is non-reversible \cite{Elfring:2009},
which implies that the time-reversed dynamics should not be equivalent to the original dynamics.
For our model swimmer with $m_1=-m_2$, 
the phase dynamics becomes reversible if $h=0$.
As a proof, consider again the rotation that
rotates the swimmer in the $xy$-plane by an angle $\pi$ around the point $(\s_1+\s_2)/2$.
The rotated swimmer is equivalent to a swimmer with
$m_1\rightarrow m_2=-m_1$, $m_2\rightarrow m_1=-m_2$ and $h\rightarrow -h$.
On the other hand, a mere rotation of the swimmer does not change its synchronization behavior.
We conclude that the phase dynamics of the original swimmer is equivalent
to the time-reversed dynamics of a swimmer with $h\rightarrow -h$.
For $h=0$, the phase dynamics is thus reversible and
all orbits of the phase dynamics are neutrally stable. 
In particular, $\lambda=0$ for $h=0$.
As a corollary, the synchronization strength $\lambda$ also vanishes
in the absence of the third sphere, since $\lambda$ is independent of $h$ in this case.
The situation is different for a swimmer that cannot translate, but still rotate around $\r_3$.
The very fact that the swimmer is acted upon by constraining forces 
to ensure $\dot{\r}_3=0$ breaks reversibility
and results in stable synchronization even in the absence of the third sphere,
provided $\omega_0h<0$, see eq.~(\ref{eq_lambda_rot}).

\section{Discussion}

In this manuscript, 
we presented a simple model swimmer to demonstrate a generic mechanism for hydrodynamic synchronization.
This synchronization mechanism depends on the bidirectional coupling between 
the phase dynamics of two oscillators that propel the swimmer and the resultant swimming motion.
If the two oscillating spheres are out-of-phase, the swimmer rotates and moves side-ways,
which imparts different hydrodynamic friction forces on these two spheres.
Since the swimming motion is determined by the orbiting of \emph{both} spheres,
this effectively couples their phase dynamics.
For free swimming, 
this synchronization mechanism is independent of direct hydrodynamic interactions between the spheres \cite{Friedrich:2012c}. 
Direct hydrodynamic interactions between the two driven spheres 
represents a second, albeit weaker, synchronization mechanism.
Restraining the swimmer's ability to move reduces the synchronization strength
in a characteristic manner, resulting in different scaling behaviors of the
synchronization strength for different constraining conditions.
For a fully clamped swimmer, synchronization arises solely from hydrodynamic interactions and is strongly attenuated.

From our analysis, we can furthermore deduce the swimmer design that is optimal for synchronization.
The existence of an optimal design can be intuitively understood as follows:
In the absence of the third sphere (which represents a passive drag center), 
the swimmer attains an additional symmetry, which rules out synchronization.
Generally, 
synchronization requires non-reversibility of the phase dynamics \cite{Elfring:2009},
which can be related to broken symmetries of the design of the swimmer.
If the third sphere is, however, very large in comparison to the two actively driven spheres,
the swimming motion of the swimmer would attenuate, and thus the synchronization strength be reduced.
In fact, we show in a limit of small spheres
that synchronization works best if all three spheres are of same size ($b=a$).
A similar reasoning applies also to the \textit{position} of the third sphere.
Synchronization is ruled out by symmetry in a swimmer design 
for which the third sphere is collinear with the rotation centers of the two driven spheres.
For a large distance between the third sphere and these rotation centers, however,
rotations of the swimmer are significantly reduced, 
which in turn implies a reduced synchronization strength.
Correspondingly, we find that the synchronization strength is maximal 
for an intermediate value of this distance ($h=l$).

Our three-sphere swimmer characterizes a novel synchronization mechanism in a minimal setting: 
the mutual interplay of swimming motion and phase dynamics 
results in a strong coupling between the two driven spheres. 
This mechanism is generic and should apply also to biological swimmers,
such as the unicellular green algae \textit{Chlamydomonas} that swims with two flagella like a breast-swimmer.
The two flagella of \textit{Chlamydomonas} can synchronize in-phase and beat in a mirror-symmetric fashion \cite{Ruffer:1985a,Polin:2009}.
We predict that asynchronous beating results in rotational motion of the cell,
which imparts different hydrodynamic forces on the two flagella, which can couple their beat.
The mere existence of a synchronization coupling between the flagella of \textit{Chlamydomonas}, however, does not tell whether this coupling stabilizes the in-phase or some out-of-phase synchronized state. 
These two cases are discriminated by the sign of the synchronization strength, 
which in turn depends on the geometry of the swimmer in a non-generic way. 
A definite answer on the sign of the synchronization strength for \textit{Chlamydomonas} 
requires detailed hydrodynamic computations using a realistic beat pattern and cell geometry. 
In a previous publication \cite{Friedrich:2012c}, 
we estimated positive values for the synchronization strength 
(indicating in-phase synchronization)
for a swimming \textit{Chlamydomonas} cell using a flagellar beat pattern tracked from experiments.
While realistic models are required for quantitative statements including the prediction of the correct sign, 
minimal models such as the three-sphere swimmer enable us to dissect the fundamental mechanisms underlying synchronization.

In addition to synchronization, our model swimmer allows us to illustrate
generic features of self-propulsion at low Reynolds numbers.
For example, our swimmer displays characteristic back-{\&}-forth wiggling on top of net propulsion,
which represents a hallmark of self-propulsion at low Reynolds numbers by cyclic swimming strokes.
Unlike net propulsion, this periodic motion requires neither
non-reversible swimming strokes nor hydrodynamic interactions.
This periodic motion is crucial for the synchronization mechanism studied here.
A wiggling back-{\&}-forth motion is not only predicted for model swimmers,
but actually observed for biological micro-swimmers that use cyclic swimming strokes.
For example, sperm cells display a characteristic wiggling motion of the sperm head,
being propelled by undulatory bending waves of their eukaryotic flagellum \cite{Gray:1955a,Friedrich:2010}.
Another class of biological swimmers, however, such as \textit{Paramecium} 
generate slip flows at their boundary to swim at a steady velocity without apparent body changes \cite{Brennen:1977}.
Our simple three-sphere swimmer incorporates both propulsion primitives as limit cases.

In conclusion, we have shown that
swimming and synchronization can strongly depend on each other.
Both net propulsion and net synchronization are a generic consequence of broken symmetries. 
The actual sign of the swimming speed or synchronization strength, however,
will depend on the geometry of the swimmer in a non-generic way.


\appendix



\section{Rotne-Prager approximation of hydrodynamic interactions} 
\label{Rod_Prag_appr}

In this appendix, we show how the grand hydrodynamic friction matrix $\G_0$,
can be computed to leading order approximation 
for an ensemble of $N$ spheres of \emph{different} radii, 
see \cite{Reichert:phd} for the case of equally-sized spheres.
The inverse of $\G_0$, the grand mobility matrix $\M_0=\G_0^{-1}$, 
describes the linear dependence between 
the translational and rotational velocities $\v_i$ and $\omegabf_i$ of the spheres, $i=1,\ldots,N$,
and the forces $\F_i$ and torques $\T_i$ exerted by the spheres on the fluid, 
\begin{align}
& \v_i = \sum_{j=1}^N \mubf^{tt}_{ij} \F_j + \mubf^{tr}_{ij} \T_j, \label{stokeslin1} \\
& \omegabf_i = \sum_{j=1}^N \mubf^{rt}_{ij} \F_j + \mubf^{rr}_{ij} \T_j \label{stokeslin2}.
\end{align}
Here, we have decomposed $\M_0$ into $3\times 3$ blocks as 
$\M_0=(\mubf_{ij}^{tt},\mubf_{ij}^{tr};\mubf_{ij}^{rt},\mubf_{ij}^{rr})_{i,j=1,\ldots,N}$
to show the different types of couplings explicitly.
With the compact notation from section \ref{sec_hydro}, we could equivalently write $\dot{\q}_0=\M_0\cdot\P_0$.

The method of reflections can be used to construct 
approximations of arbitrary precision of the mobility matrix $\M_0$ \cite{Dhont:1996,Reichert:phd}.
The first iteration step of this method accounts for all pair-wise interactions between spheres
and is known as the Rotne-Prager approximation.
Specifically, we treat a sphere with index $j$ as a test particle that is convected
by the fluid flow induced by another sphere with index $i$ that moves
under the influence of a force (or torque).
Higher-order corrections would for example describe how the presence of the $j$th sphere
changes the mobility of the $i$th sphere.

We denote the sphere's positions by $\r_i$ and their radii by $a_i$, $i=1,\ldots,N$.
The flow field $\u_i^t$ induced by a single sphere of radius $a_i$ 
translating under the influence of an external force $\F_i$ with a constant velocity 
$\v_i=\F_i/\gamma_i$, $\gamma_i=6\pi\eta a_i$,
in a viscous fluid of viscosity $\eta$ reads
\begin{equation}
\label{eq_ut}
  \u_i^t(\r,t) = A(\r-\r_i) \v_i, \quad
         A(\r) =   \frac{3}{4} \frac{a_i}{r} (\mathbf{1}+\hat{\r}\otimes\hat{\r})
                 + \frac{1}{4}\left(\frac{a_i}{r}\right)^3(\mathbf{1}-3\hat{\r}\otimes\hat{\r}).
\end{equation}
Here $r=|\r-\r_i|$, $\hat{\r} = (\r-\r_i)/r$.
If the sphere is subject to an external torque $\T_i$ and rotates
with rotational velocity $\omegabf_i=\T_i/\gamma_i^r$, $\gamma_i^r=8\pi\eta a_i^3$,
then the induced flow field is
\begin{equation}
\label{eq_ur}
\u_i^r(\r)=\left(\frac{a_i}{r}\right)^3 \omegabf \times \r.
\end{equation}

Faxen's theorem provides the translational and rotational velocities of a second sphere with label $j$
that is free from external forces and torques and becomes convected by the flow $\u_i$
\begin{align}
& \v_j = \mathcal{L}^t \u(\r)|_{\r=\r_j},\quad \mathcal{L}^t=1+\frac{1}{6}a_j^2 \nabla^2, \\
& \omegabf_j = \mathcal{L}^r \u(\r)|_{\r=\r_j},\quad \mathcal{L}^r=\frac{1}{2} \nabla \! \times.
\end{align}
Combining the explicit expressions for the flow fields induced by the $i$th sphere 
given by eqns.~(\ref{eq_ut}),(\ref{eq_ur}) and Faxen's laws provides 
leading order approximations for the cross-mobilities $\mubf_{ji}$ \cite{Dhont:1996,Reichert:phd}
\begin{align}
\label{eq_rp_tt}
  &\mubf^{tt}_{ji} 
    =    \frac{1}{8\pi\eta r_{ij}}(\mathbf{1} + \hat{\r}_{ij}\otimes\hat{\r}_{ij})
       + \frac{a_i^2+a_j^2}{24\pi\eta r_{ij}^3} (\mathbf{1}-3\hat{\r}_{ij}\otimes\hat{\r}_{ij})&
      &+\O(\eps^4)& \\
  &\mubf^{rr}_{ji}
    =  - \frac{1}{16\pi\eta r_{ij}^3}(\mathbf{1}-3\hat{\r}_{ij}\otimes\hat{\r}_{ij})&
      &+\O(\eps^6)& \\
  &\mubf^{rt}_{ji}=\mubf^{tr}_{ji} 
    =    \frac{1}{8\pi\eta r_{ji}^2} \hat{\r}_{ij} \times &
      &+\O(\eps^5). &
\end{align}
Here, we introduced a small expansion parameter $\eps$ and assumed that all sphere radii 
are small of order $\eps$ compared to the distances $r_{ij}=|\r_{ij}|$ with $\r_{ij}=\r_j-\r_i$.
The notation $\hat{\r}_{ji}\times$ stands for the rank-2-tensor with components $(\hat{\r}_{ji}\times)_{lm}=\eps_{klm} (\hat{\r}_{ji})_k$,
where $\eps_{klm}$ is the Levi-Civita tensor.
The self-mobilities are perturbed only to higher order \cite{Dhont:1996,Reichert:phd}
\begin{equation}
 \mubf^{tt}_{ii} = \frac{1}{\gamma_i^t} \mathbf{1} + \O(\eps^4), \quad
 \mubf^{rr}_{ii} = \frac{1}{\gamma_i^r} \mathbf{1} + \O(\eps^6), \quad
 \mubf_{ii}^{rt} = \O(\eps^7),
\end{equation}
The analytical results presented in the main text do not depend on higher corrections of the mobility matrix.

\section{A limit of small spheres and lever arms} 
\label{perturb_calc}

To gain analytical insight into the equation of motion, 
we resort to a limit of a small beat amplitude characterized by small lever arms and small spheres.
We introduce a small expansion parameter 
$\eps=a/l$ and assume both $b/l$ and $R/l$ to be of order $\O(\eps)$. 
We further assume that $\kappa/(\eta l^3)$ is of order unity.
We expand the force balance given by eqn.~(\ref{eq_balance}) as a power series in $\eps$
making use of the block matrix decomposition (\ref{gamma_blocks})
\begin{align*}
 \begin{pmatrix} 0\\ 0\\ 0\\ m_1\\ m_2 \end{pmatrix}
=
 \left( \begin{tabular}{rlrrrrr}
       $\eps \K^{(1)}$      &$+ \eps^2 \K^{(2)}$       &$+ \ldots$ & \;&
       $\eps^2 \C^{(2)}$    &$+ \eps^3 \C^{(3)}$       &$+ \ldots$ \\
       $\eps^2 \C^{(2),T}$  &$+ \eps^3 \C^{(3),T}$     &$+ \ldots$ & \;&
       $\Omegabf^{(0)}$     &$+ \eps^3 \Omegabf^{(3)}$ &$+ \ldots$ 
    \end{tabular} \right)
 \left( \begin{tabular}{rll}
       $\eps \dot{\X}^{(1)}$      &$+ \eps^2 \dot{\X}^{(2)}$      &$+ \ldots$ \\
       $\dot{\Phibf}^{(0)}$       &$+ \eps^3 \dot{\Phibf}^{(3)}$   &$+ \ldots$ 
    \end{tabular} \right).
\end{align*}
Note that $\K^{(1)}$ and $\Omegabf^{(0)}$ are invertible.
We can now iteratively solve for the series coefficients of $\dot{\X}$ and $\dot{\Phibf}$.
Hydrodynamic interactions between the three spheres contribute to leading order 
only to $\K^{(2)}$, $\C^{(3)}$ and $\Omegabf^{(4)}$.

\def\urlprefix{}
\def\url#1{}
\bibliographystyle{apsrev}
\bibliography{C:/ben/bibliography/library}

\newpage


\begin{spacing}{1.0}

\begin{figure}
\begin{center}
\includegraphics[width=15cm]{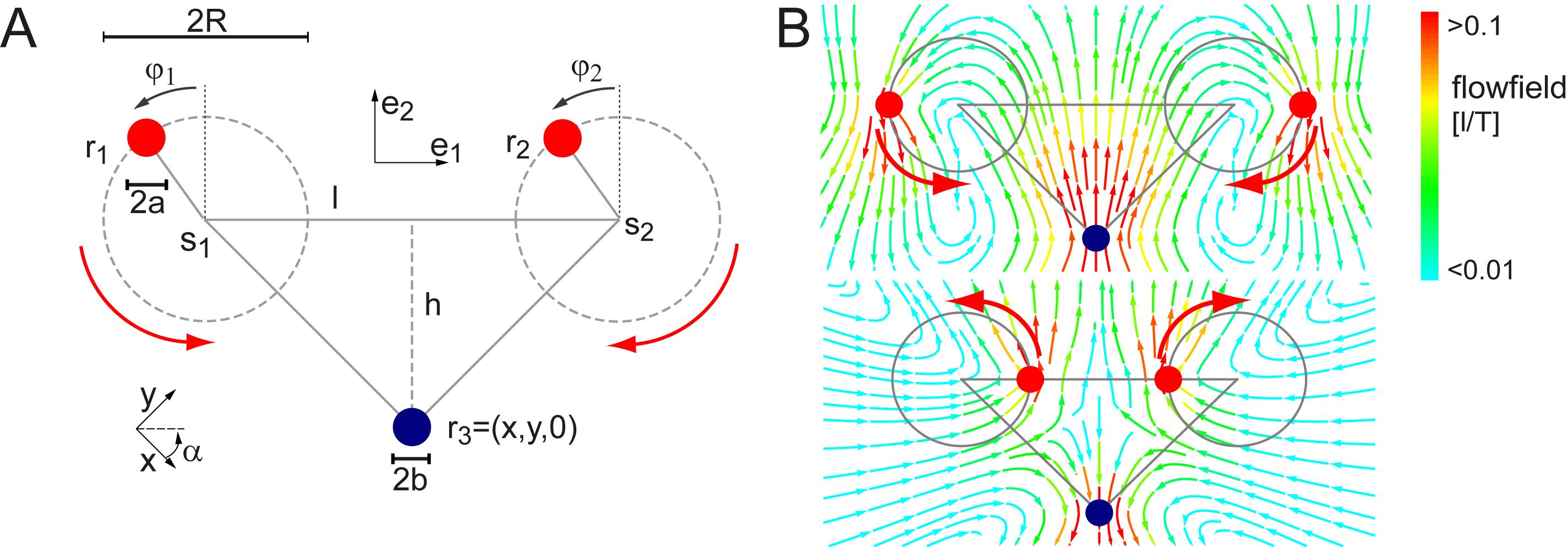}
\end{center}
\caption[]{
A simple three-sphere swimmer.
\textbf{A.} 
The swimmer consists of three spheres connected by a frictionless, planar scaffold.
The first and second sphere (red, located at $\r_1$ and $\r_2$) can move along
circular orbits, thus propelling the swimmer.
The swimmer is characterized by 5 degrees of freedom:
two phase angles $\varphi_1$ and $\varphi_2$ that parametrize the circular orbits of the first and second sphere,
as well as the planar position $(x,y)$ and orientation angle $\alpha$ of the third sphere (blue, located at $\r_3$).
\textbf{B.}
Flow fields induced by the swimming motion of the swimmer 
when immersed in a viscous fluid.
Parameters: $a=b=0.1\,l$, $h=2R=l$, $m_1=\kappa=\eta l^3$.
}
\end{figure}

\begin{figure}
\begin{center}
\includegraphics[width=15cm]{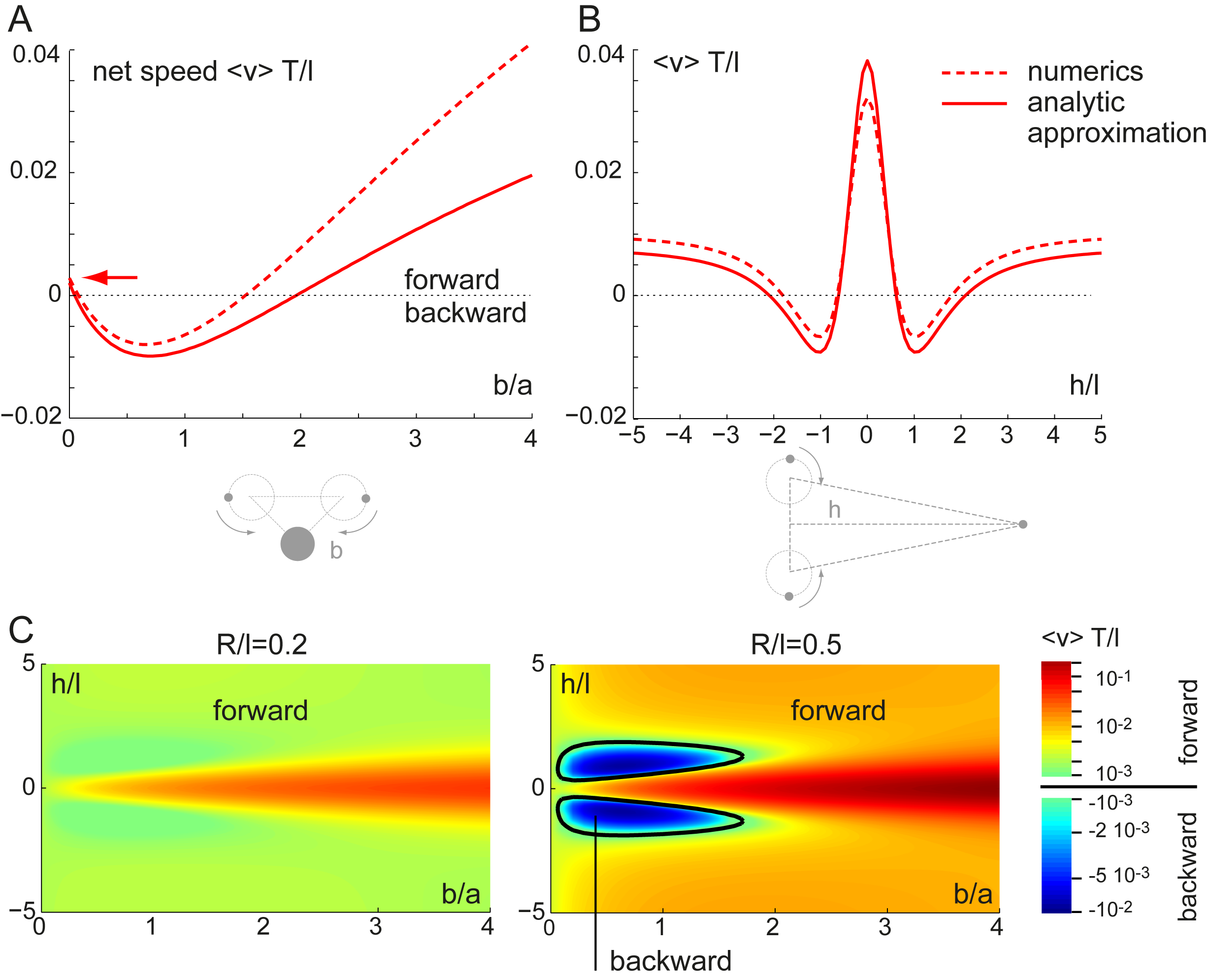}
\end{center}
\caption[]{
Forward or backward swimming for different swimmer geometries.
For synchronized phase dynamics with $\delta=\varphi_1+\varphi_2=0$,
the swimmer will move along a straight path with velocity $v=\r_3\cdot\e_2$.
\textbf{A.}
The net swimming velocity, averaged over a beat cycle, 
depends on the radius $b$ of the third sphere (\textbf{A})
as well as on its position $h$ (\textbf{B}).
Solid lines indicate the analytical approximation,
while the dashed line was determined from numerical integration of the equation of motion.
\textbf{C.}
Numerical results for the net swimming speed as a function
of both $b$ and $h$ for two different values of the lever arm length $R$.
Parameters: As in figure 1, unless indicated otherwise.
}
\end{figure}

\begin{figure}
\begin{center}
\includegraphics[width=15cm]{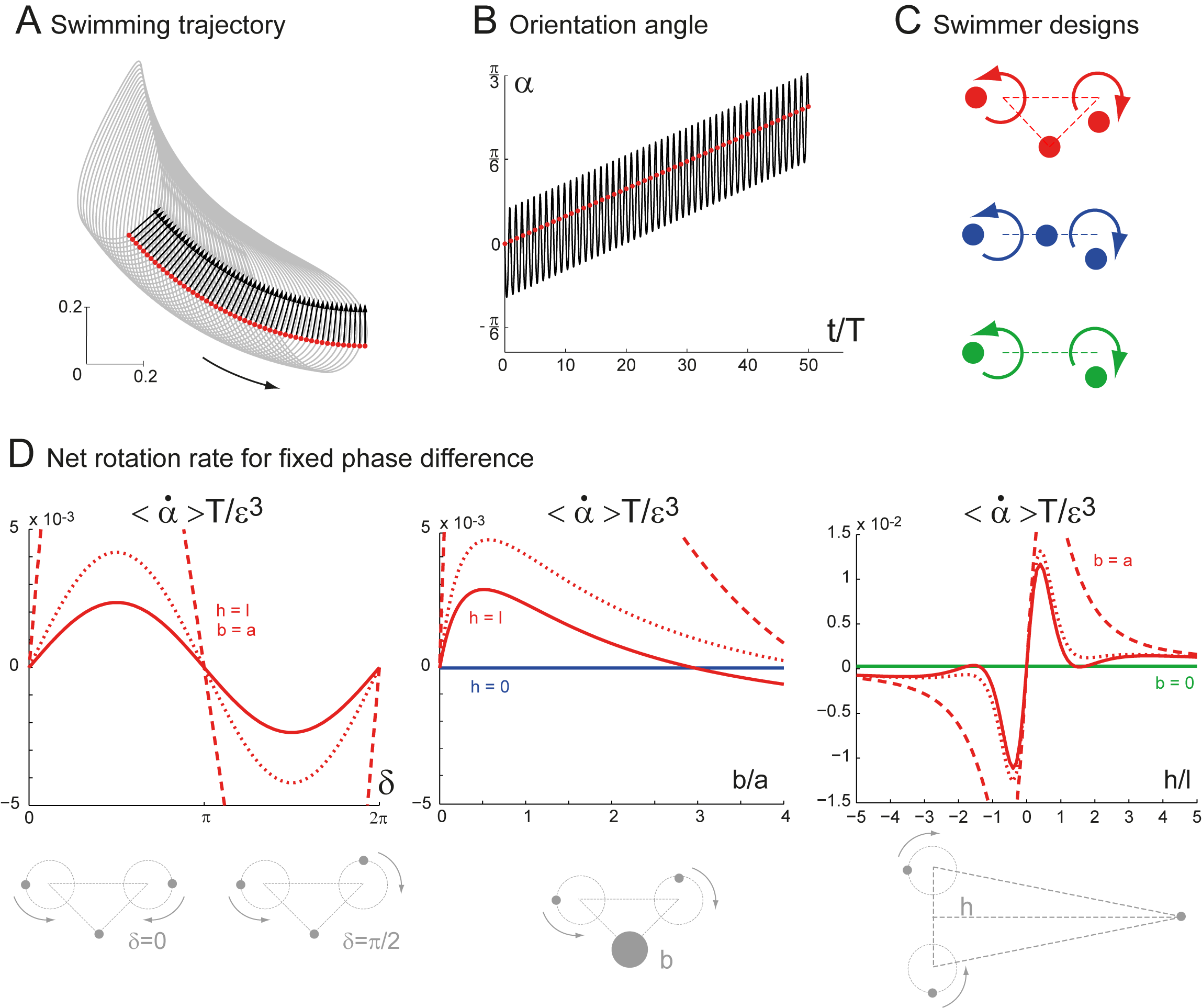}
\end{center}
\caption[]{
Asynchronous beating causes rotation.
For this figure only, we us a prescribed phase dynamics
with $\varphi_1=\omega_0 t$ and $\varphi_2=-\omega_0 t+\delta$
to separate the swimming motion from its feedback on the phase dynamics.
\textbf{A.}
Swimming trajectory $\r_3(t)$ for constant phase difference $\delta=\pi/2$.
The position $\r_3(t_n)$ and orientation $\e_2(t_n)$
after each full cycle of the first sphere, $\varphi_1(t_n)=2\pi n$, 
is indicated as dot and arrow, respectively.
\textbf{B.}
The orientation angle $\alpha=\alpha_3(t)$ oscillates 
around a linear trend, indicating non-zero net rotation after each full cycle.
\textbf{C.}
Different swimmer designs studied in panel D:
top (red): the design from figure 1,
middle (blue): $h=0$,
bottom (green): without the third sphere, $b=0$.
\textbf{D.}
For the swimmer design from figure 1, 
the net rotation rate $\langle\dot{\alpha}\rangle$
depends on the phase difference $\delta$
as well as on the radius $b$ and position $h$ of the third sphere
(red solid line: analytical approximation, 
red dashed: numerical result).
Additionally, we show numerical results for the case of
scaled-down spheres and lever arms with 
$a=b=\eps\,l$, $R=5\eps\,l$ using $\eps=0.01$ (red dotted curve).
The net rotation rate vanishes due to symmetry reasons
for $b=0$ (blue curve) and $h=0$ (green curve).
Parameters: As in figure 1-2, unless indicated otherwise;
panel D, middle and right: $\delta=\pi/2$.
}
\end{figure}

\begin{figure}
\begin{center}
\includegraphics[width=15cm]{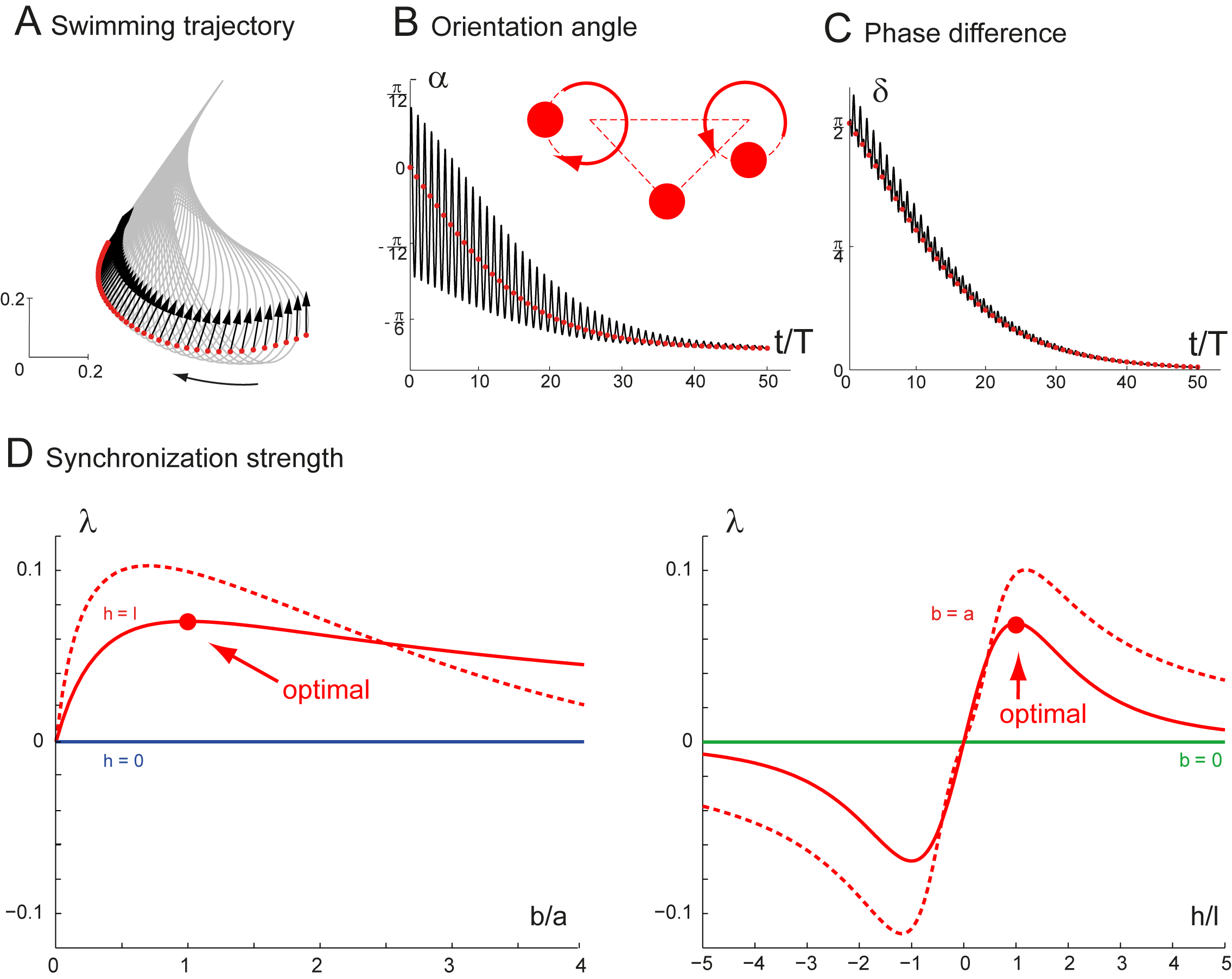}
\end{center}
\caption[]{
Phase-synchronization of a free-moving swimmer.
\textbf{A.}
Swimming trajectory $\r_3(t)$ for a swimmer with $\omega_0<0$
and initial phase difference $\delta(0)=\pi/2$.
The position $\r_3(t_n)$ and orientation $\e_2(t_n)$
after each full cycle of the first sphere, $\varphi_1(t_n)=-2\pi n$, 
is indicated as dot and arrow, respectively.
\textbf{B.}
The orientation angle $\alpha=\alpha_3(t)$ 
oscillates with an amplitude that decays in time as the two driven spheres synchronize in-phase.
\textbf{C.}
The phase-difference $\delta=\varphi_1+\varphi_2$ decays towards zero,
indicating in-phase synchronization.
\textbf{D.}
Stable in-phase synchronization is characterized by a synchronization strength $\lambda>0$
(solid line: analytical approximation eq.~(\ref{eq_lambda}), dashed line: numerical result).
Here, $\lambda$ is defined as $\lambda=-d\Lambda(\delta)/d\delta_{|\delta=0}$,
where $\Lambda(\delta(t_0))=\delta(t_1)-\delta(t_0)$ measures the change in 
phase difference after one full cycle of the first sphere.
In a limit of small spheres, $\eps\ll 1$,
synchronization strength $\lambda$ is maximal for a swimmer design with $b=a$, $h=l$ as indicated.
In the absence of the third sphere, $b=0$, the synchronization strength vanishes,
since the dynamics of the swimmer becomes reversible in this case.
Similarly, one can derive the symmetry relation 
$\ln[1+\lambda(-h)]=-\ln[1+\lambda(h)]$,
which implies that the in-phase synchronized state with $\delta=0$
changes its stability if $h$ reverses sign.
Parameters: As in figure 1-3, except that the sense of rotation
of the first and second sphere is reversed, $-m_1=m_2=\kappa=\eta l^3$.
}
\end{figure}

\end{spacing}

\end{document}